\documentclass[pre,superscriptaddress,twocolumn]{revtex4}
\usepackage{epsfig}
\usepackage{latexsym}
\usepackage{amsmath}
\begin {document}
\title {Splitting the voter criticality}
\author{Michel Droz}
\affiliation{Department of Physics, University of Geneva, CH 1211
Geneva 4, Switzerland}
\author{Antonio L.~Ferreira}
\affiliation{Departamento de Fisica, Universidade de Aveiro, 3810-193
 Aveiro, Portugal}
\author{Adam Lipowski}
\affiliation{Department of Physics, University of Geneva, CH 1211
Geneva 4, Switzerland}
\affiliation{Faculty of Physics, A.~Mickiewicz University,
61-614 Pozna\'{n}, Poland}
\begin {abstract}
Recently some two-dimensional models with double symmetric absorbing 
states were shown to share the same critical behaviour that was called the 
voter universality class.
We show, that for an absorbing-states Potts model with finite but further 
than nearest neighbour range of 
interactions the critical point is splitted into two critical points: one of 
the Ising type, and the other of the directed percolation universality class.
Similar splitting takes place in the three-dimensional nearest-neighbour model.
\end{abstract}
\maketitle
Nonequilibrium phase transitions are recently attracting increasing theoretical 
interest.
One of the motivations driving the research in this field is the idea that, 
similarly to equilibrium systems, nonequilibrium continuous phase transitions 
can be 
also divided into several, hopefully not so many, universality 
classes~\cite{HAYE1}.
Indeed, there are some examples that show that such a classification, at 
least to some extent, can be made.
Particularly interesting results exist for models with absorbing states, and a 
notable example is the directed percolation (DP) universality class.
A conjecture by Jenssen and Grassberger~\cite{DP} that models 
with a single 
absorbing state should belong to the same universality class has by now 
very convincing supports.
Another universality class encompasses models with double absorbing 
states~\cite{PC1,HAYE97}, that in one dimension ($d=1$) includes also models 
which dynamics has some additional symmetry (parity conservation)~\cite{PC}.
However, it is also known that any asymmetry in the dynamics that would favor 
any of the absorbing states drives the system into DP universality 
class~\cite{PARK}.
Universality of models with more than two absorbing states is even more 
problematic since for models with symmetric absorbing states some details of 
the dynamics might affect the critical behaviour too~\cite{LIPDROZ1}.
It is becoming  clear that a task of classifying nonequilibrium phase 
transitions is far from completed~\cite{ODOR}.

Numerical results that support the above classification come mainly from 
one dimensional models.
The situation for $d>1$ is less understood, but there are some
results in this case too.
In addition to numerous examples of models with a single absorbing state and 
thus belonging to DP universality class~\cite{HAYE1}, there are some indications that
for $d>1$ models with more than
two symmetric absorbing states the phase transition should be 
discontinuous~\cite{LIPDROZ}.
In between, there are models with two absorbing states.
Dornic et al. shown that in this case and for $d=2$ a group of models
belongs to a new universality class called a voter universality 
class~\cite{CHATE}.
Originally, a voter model was introduced in a rather nonphysical 
context of opinion spreading~\cite{LIGGETT}.
Later on numerous examples of related and physically more relevant models were 
also studied~\cite{VOTER}.
Although the phase transition in the voter universality class is continuous, 
the decay of the order 
parameter $\rho$ upon approaching the critical point is slower than any  
power law decay $\rho\sim \epsilon^{\beta}$, where $\epsilon$ measures the
distance from the critical point.
Formally such a decay might be described with $\beta=0$.
In addition, the time decay of the order parameter at criticality is also 
slower than any power law decay and is in fact logarithmic 
($\rho\sim 1/{\rm ln}(t)$) as shown exactly~\cite{KRAP}.
Some other exponents of voter universality class were also 
determined~\cite{CHATE}.

An interesting feature of the voter model, that
was not yet addressed, is the fact that at its critical point actually two 
phenomena seem to take place.
One of them is the symmetry breaking between two competing states of the model,
that is similar to the symmetry breaking in the 
equilibrium Ising model.
The second phenomenon is the phase transition between active and absorbing 
phases of the model.
Since there is no symmetry breaking transition in one dimensional equilibrium 
Ising model, it is easy to understand that for models with two
absorbing states in $d=1$,
the symmetry breaking must result from the absorbing phase transition and thus
both phenomena take place simultaneously.
But this is no longer the case for $d>1$ models, and the coincidence of 
these two transitions should not be taken for granted.

In the present paper we study a recently introduced nonequilibrium
Potts model which dynamics has two absorbing states~\cite{LIPDROZ}.
We show that the model with nearest-neighbour interactions on square 
lattice belongs to the voter universality class.
However, with extended range of interactions (up to 
third nearest neighbours) the voter critical point is splitted.
Starting from the disordered phase and reducing a temperature-like control 
parameter $T$, the model first undergoes symmetry breaking phase transition.
Calculation of the Binder cumulant indicates that this transition belongs to 
the Ising universality class.
Upon further decrease of $T$, the model undergoes a second phase transition 
into the absorbing phase.
Since at this point the symmetry is already broken, this second transition, as
expected, belongs to the DP universality class.
Similar behaviour is observed for the nearest neighbour model in the 
three-dimensional case.

That a certain type of a nonequilibrium critical point can be considered as 
superposition of two other critical points is in our opinion new, and 
hopefully, it should increase the understanding of nonequilibrium phase 
transitions.
Let us notice that a superposition of different critical points 
exist in some equilibrium systems.
For example, in the frustrated XY model two phase transitions of the
Kosterlitz-Thouless type ($U(1)$) and of the Ising type ($Z_2$) under
certain conditions most likely happen simultaneously~\cite{XY}.
Another example is a multicritical point in random magnets~\cite{AHARONY}.

Before defining our model let us recall that the equilibrium 
ferromagnetic 2-state Potts model can be defined using the following 
Hamiltonian~\cite{WU}
\begin{equation}
H= -\sum_{(i,j)} \delta_{\sigma_i\sigma_j},
\label{e1}
\end{equation}
where summation is over pairs $(i,j)$ of interacting sites on a cartesian 
lattice of the linear size $L$.
With each site $i$ we assign a variable $\sigma_i=0,1$ and $\delta$ is the 
Kronecker delta function.

To study model~(\ref{e1}) using Monte Carlo simulations one
constructs a stochastic Markov process with suitably chosen transition
rates.  
One of the possible choices corresponds to the so-called
Metropolis algorithm.  
In this algorithm one looks at the energy
difference $\Delta E$ between the final and initial configuration and
accept the move with probability min$\{1,{\rm e}^{-\Delta E/T}\}$,
where $T$ is temperature.
To obtain a model that would have symmetric absorbing states we modify the 
Metropolis 
algorithm of model~(\ref{e1}) as follows~\cite{LIPDROZ}: when all neighbours
of a given site $i$ are in the same state as this site, then the site $i$ 
cannot change its state.
Thus, the dynamics of our model is defined as follows:
(i) Select randomly the site $i$ and its possible final state.
(ii) If $\Delta E<z$ update the site $i$ with the probability 
min$\{1,{\rm e}^{-\Delta E/T}\}$, where $z$ is the number of neighbours 
interacting with site $i$ (in our paper $z$ is $i$-independent).
Let us notice, that after such a modification, $T$ is no longer temperature.
Nevertheless, we will keep such a terminology.
Moreover, the unit of time is defined as a single (on average) update of every
site.

To examine the properties of our model we used numerical 
simulations. 
A natural characteristic of models with absorbing states
is the steady-state density of active sites $\rho$.
In our model a given site $i$ is active when at least one of its 
neighbours is in a state different than $i$.
Otherwise the site $i$ is called nonactive.
Moreover, we used the so-called dynamic Monte Carlo method where one
sets the system in the absorbing state with activity only locally
initiated and measure some stochastic properties of runs~\cite{GRASSTORRE}.  
Typical characteristics are the survival probability $P(t)$ that activity
survives at least until time $t$, and the average number of active sites $N(t)$
(averaged over all runs).
At criticality $P(t)$ and $N(t)$ are expected to
have power-law decay: $P(t)\sim t^{-\delta}$ and $N(t)\sim t^{\eta}$.
To detect a possible symmetry breaking in the model we measured the
magnetization $m=\frac{2}{L^d}\langle\sum_i \sigma_i\rangle-1$ and its 
variance 
$s=\frac{1}{L^d}\langle(2\sum_i (\sigma_i-\langle\sigma_i\rangle) -L^d)^2\rangle$.
At the symmetry breaking critical point the variance $s$, that is related 
with magnetic 
susceptibility for equilibrium systems, is expected to diverge in the limit 
$L\rightarrow \infty$.
Below we present the results of our simulations.
We ensured that the lattice size $L$ is large enough for the finite-size 
effects to be negligible.

\begin{figure}
\centerline{\epsfxsize=8cm
\epsfbox{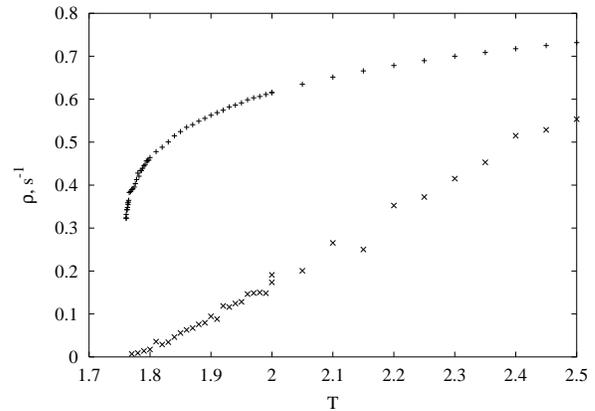}
}
\caption{
The density of active sites $\rho$ (+) and the inverse
of the variance of magnetization $s$ multiplied by a factor of 5 ($\times$) as 
a function of temperature for the two dimensional nearest neighbour 
model.
Simulations were done for $L=500$.
Close to the critical temperature $T_c$(=1.7585~\cite{LIPDROZ}) the linear decay
of $s^{-1}$ might be affected by some logarithmic corrections~\cite{CHATE}.
}
\label{magnn}
\end{figure}

\noindent (A) Model with nearest-neighbour interactions on square lattice\\
In this case the model was already studied by some of us, but only in the 
context of an absorbing phase transition~\cite{LIPDROZ}.
A very slow decay of $\rho$ at the critical point was observed, but 
being unaware of the relation with the voter model, we could not
draw definite conclusions about the nature of the critical point in this 
case~\cite{CHATE1}.
Indeed, having a double symmetric absorbing state and being driven only by interfacial 
noise, the model satisfies the Dornic et al.'s criterions for 
belonging to the voter universality class.
Additional confirmation is shown in Fig.~\ref{magnn}.
One can see that the variance $s$ diverges at the same temperature where $\rho$
vanishes.
Moreover, $s^{-1}$ seems to decay linearly at the critical point which implies
that $\gamma=1$, as also predicted for this universality class~\cite{CHATE}.

\noindent (B) Model with up to third-nearest-neighbour interactions on 
square lattice\\
Studying our nonequilibrium Potts model within a mean-field approximation at 
the pair level, we  noticed that for sufficiently large
coordination number $z$ a structure of the solution qualitatively changes in 
a way
that clearly indicates two separate transitions in the model~\cite{PREPARATION}.
It prompted us to simulate our model with interactions including also further 
neighbours.
First we studied the model with interactions up to second-nearest neighbours 
($z=8$).
It turned out, that in this case either the model belongs to the voter
universality class or there is only extremely small splitting beyond the 
resolution of our simulations.

\begin{figure}
\centerline{\epsfxsize=8cm
\epsfbox{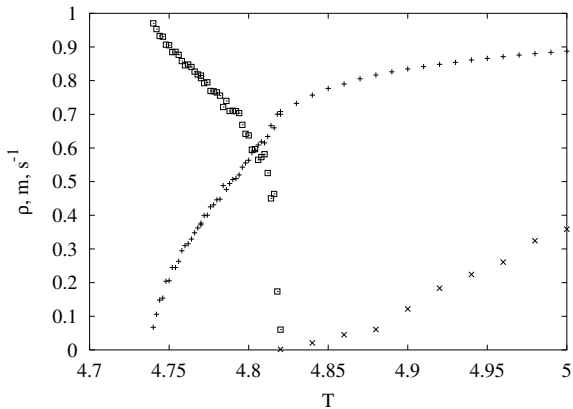}
}
\caption{
The density of active sites $\rho$ (+), the magnetization $m$ ($\Box$), and 
the inverse
of the variance of magnetization $s$ multiplied by a factor of 30 ($\times$) as 
a function of temperature for the two dimensional model with 12 
neighbours ($L=400$).}
\label{mag}
\end{figure}

However, for the model with interactions up to third-nearest 
neighbours ($z=12$), a qualitatively new picture emerges.
Indeed, one can see in Fig.~\ref{mag} that the variance $s$ diverges at the 
temperature $T_I$ that is clearly larger than the temperature $T_c$ where
$\rho$ vanishes.
In the temperature interval $T_c<T<T_I$ our model is magnetized 
(Fig.~\ref{mag}).
To avoid rather slow coarsening effects in this interval it is better to start
simulations from an asymmetric ($m\neq 0$) initial configuration.

\begin{figure}
\centerline{\epsfxsize=7cm
\epsfbox{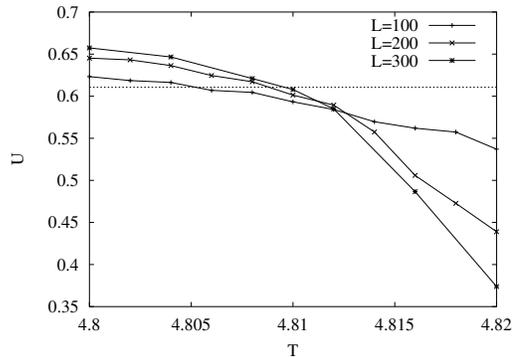}
}
\caption{
The Binder cumulant $U$ as a function of temperature $T$ for the two 
dimensional model with 12 neighbours.
The horizontal dotted line denotes the universal value $U\sim 0.6107$ for the
$d=2$ Ising model.
To diminish fluctuations long simulations were made with the 
simulation time $t\sim 5\cdot 10^6$ Monte Carlo steps.
}
\label{binder}
\end{figure}

To examine the nature of the phase transition at $T=T_I$, we calculated the 
so-called Binder cumulant~\cite{BINDER} $U=1-\frac{m_4}{3m_2^2}$, where
$m_n$ is the $n$-th moment of magnetization.
In Fig.~\ref{binder} one can see that at the crossing point $U$ is relatively 
close to the universal value $U=0.6107$ of the $d=2$ Ising 
model~\cite{KAMIENIARZ}.
Although our model is nonequilibrium its critical behaviour at $T=T_I$ is 
the same as in equilibrium (Ising) systems.
Such a feature is in agreement with some expectations~\cite{GRINSTEIN} that at
the critical point of many nonequilibrium systems only some general
properties (e.g., symmetry) determines the nature of the critical point 
while some 
other, as e.g., lack of detailed balance, are very often irrelevant in this 
respect.
\begin{figure}
\centerline{\epsfxsize=7cm
\epsfbox{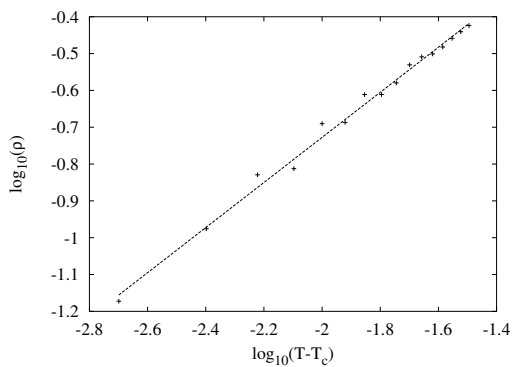}
}
\caption{
The scaling of the density of active sites $\rho$ in the vicinity of the
critical point $T_c=4.738$ in the two dimensional model with 12 
neighbours.
Simulations were done for $L=500$ and the linear fit has a slope $\beta=0.61$.}
\label{maglog}
\end{figure}
\begin{figure}
\centerline{\epsfxsize=8cm
\epsfbox{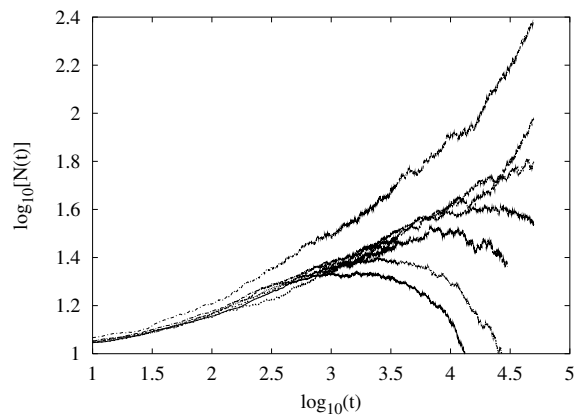}
}
\caption{
The average of number of active sites $N(t)$ as a function of time $t$
calculated using dynamical Monte Carlo for the two dimensional next-next-nearest 
neighbour model and (from top) $T=4.739$, 4.7385, 4.738 (critical point), 
4.7375, 4.737, 4.736, 4.735.
For each temperature the average is made usually over $2\cdot 10^4$ independent
runs ($L=5000$).
}
\label{dyn12a}
\end{figure}
Since the symmetry is already broken upon approaching the critical point 
at $T=T_c$, 
we expect that this critical point should belong to the (2+1) DP universality 
class.
Simulations confirm these expectations.
In Fig.~\ref{maglog} one can see that close to the critical point at $T=T_c$
the density of active sites behaves as $\rho\sim (T-T_c)^{\beta}$ and we
estimate that $\beta=0.61(4)$, that can be compared with the DP value 
0.584~\cite{HAYE1}.
Additional confirmation is obtained using the dynamical Monte Carlo method, 
that enables us also to precisely locate 
the critical temperature $T_c=4.7380(5)$.
From the measurement of $N(t)$ (Fig.~\ref{dyn12a}) at criticality we 
estimate $\eta=0.25(3)$ which is in a reasonable agreement with DP value 
0.230~\cite{HAYE1}.
We also measured $P(t)$ and from these data (not presented here) we estimate 
$\delta=0.44(3)$ that can be compared with the DP value 
0.451~\cite{HAYE1}.
Our estimation of dynamical exponents is much different from that obtained
for some models of voter universality class $\delta\sim 0.9$, 
$\eta\sim 0.0$~\cite{HAYE97,LIPDROZ}.

Let us notice that similarly to the nearest-neighbour case, 
the $z=12$ model also has two absorbing states and is driven only by 
interfacial noise.
Nevertheless, the critical behaviour in this case is much different from the
voter model.
This is thus yet another example that shows that simple criterions most likely 
cannot be used
for classifying nonequilibrium critical points.
\begin{figure}
\centerline{\epsfxsize=8cm
\epsfbox{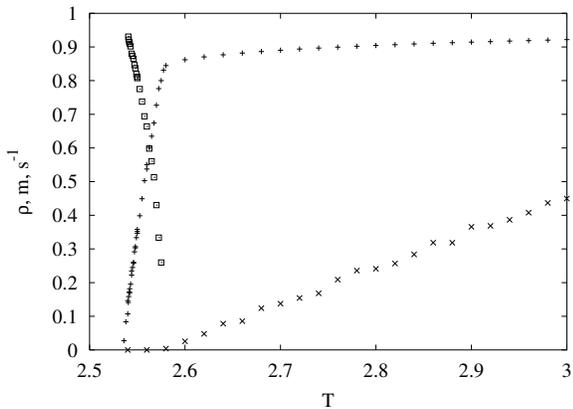}
}
\caption{
The density of active sites $\rho$ (+), the magnetization $m$ ($\Box$), and 
the inverse
of the variance of magnetization $s$ multiplied by a factor of 5 ($\times$) as 
a function of temperature for the three dimensional nearest neighbour 
model ($L=60$).}
\label{magc}
\end{figure}

\noindent (C) Model with nearest-neighbour interactions on simple cubic 
lattice\\
We also studied the three-dimensional nearest-neighbour version of our 
model (z=6).
Similarly to the $z=12$ case, here again the absorbing and magnetic phase 
transitions are separated (Fig.~\ref{magc}).
We expect that in this case the magnetic phase transition belongs
to the $d=3$ Ising universality class and the absorbing phase transition 
belongs to the (3+1) DP universality class.
Confirmation of such a scenario will require, however, extensive numerical 
simulations and is left for the future.

In summary, we have shown that models with two absorbing states in $d>1$
dimensions might exhibit two transitions where the first one breaks the 
symmetry and the second one brings the model into an
absorbing state.
In the voter model and some related models these two transitions coincide.
Hopefully, such an interpretation of the voter model will contribute to a
better understanding of its unusual critical behaviour.
For example, diminishing the strength of the further neighbour interactions
we can reduce the splitting $T_I-T_c$ and examine a crossover to the voter
universality class at which $T_I=T_c$.
Of course the Ising type phase transition is not the only type of the symmetry
breaking and other types as e.g., 3-state Potts or XY, are also possible in
nonequilibrium systems.
It would be interesting to examine whether such critical points can be 
superposed with DP universality class, that might result in new types of
critical behaviours.

We thank H.~Chat\'e for interesting discussion.
This work was partially supported by the Swiss National Science Foundation
and the project OFES 00-0578 "COSYC OF SENS".
A.~L.~F. is grateful for the financial support from
the Sapiens project POCTI/33141/99.

\end {document}